\documentclass[apj]{emulateapj}

\shorttitle{Constraints on the Magellanic Clouds' Interaction}

\begin{document}

\title{Constraints on the Magellanic Clouds' Interaction 
from the Distribution of OB Stars and the Kinematics of Giants}

\author{Dana I. Casetti-Dinescu\altaffilmark{1}, 
Katherine Vieira\altaffilmark{2},
Terrence M. Girard\altaffilmark{1} and
William F. van Altena\altaffilmark{1}}

\altaffiltext{1}{Astronomy Department, Yale University, P.O. Box 208101,
New Haven, CT 06520-8101, USA, dana.casetti@yale.edu, terry.girard@yale.edu, william.vanaltena@yale.edu}
\altaffiltext{2}{Centro de Investigaciones de Astronomia, Apartado Postal 264,
M\'{e}rida, 5101-A, Venezuela, kvieira@cida.ve}

\begin{abstract}
Young, OB-type candidates are identified in a $\sim 7900$ deg$^2$ region encompassing
the Large and Small Magellanic Clouds (LMC/SMC) periphery,
the Bridge, part of the Magellanic Stream (MS) and Leading Arm (LA). Selection is based on UV, 
optical and IR photometry from 
existing large-area surveys and proper motions from the Southern Proper Motion 4 catalog (SPM4).
The spatial distribution of these young star candidates shows  1) a
well-populated SMC wing which continues westward with two branches partially surrounding the SMC,
2) a rather narrow path from the SMC wing eastward toward the LMC which is
offset by $1\arcdeg-2\arcdeg$ from the high-density H I ridge in the Bridge,
3) a well-populated periphery of the LMC dominated by clumps of stars 
at the ends of the LMC bar and 4) 
a few scattered candidates in the MS and two overdensities in the LA
regions above and below the Galactic plane.
Additionally, a proper-motion analysis is made of a
radial-velocity selected sample of red giants and supergiants in the LMC, previously shown
to be a kinematically and chemically distinct subgroup, 
most likely captured from the SMC. SPM4 proper motions of these stars also 
indicate they are distinct from the LMC population. 
The observational results presented here, combined with the known orbits of the Clouds,
and other aspects of the LMC morphology, suggest an off-center, moderate to highly-inclined 
collision between the SMC and the LMC's disk that took place between 100 and 200 Myr ago.

\end{abstract}


\section{Introduction}

The Magellanic Clouds (MC) offer an exemplary insight into galaxy interactions, which are 
known to be an important driver of galaxy evolution.
This is due to their proximity to the Milky Way (MW) that allows for detailed 
mapping of their gaseous and stellar components, the 3D kinematics of their stellar component, and 
the chemical-abundance makeup of these components. The strongest clues to their interaction
are the H I features known as the 
Magellanic Stream (MS), the Bridge, and the Leading Arm (LA) (e.g., Nidever et al. 2010
and references therein). Earlier work attributed the formation of these structures as being 
due to the tidal interaction between the Clouds and the MW, with the understanding that the 
SMC's tidal disruption produced the MS and LA 
(e.g., Gardiner \& Noguchi 1996, Connors et al. 2006). Another scenario for the origin of the MS
is ram-pressure stripping of the LMC gas due to the motion of the Cloud through the gas in the
Galactic halo (Mastropietro et al. 2005); this model however can not explain the LA.
This earlier work was based on proper-motion measurements of small absolute value 
and the MW halo being modeled primarily as an isothermal sphere, which led to bound 
orbits of the Clouds around the MW.

More recent work offers a
somewhat different picture. The HST proper-motion measurements of the Clouds
by Kallivayalil et al. (2006a,b, hereafter K06, see also Piatek et al. 2008) 
indicate that the Clouds are on their first passage (Besla et al. 2007, 2010) about the MW. 
The MS and LA are made of material pulled out from the SMC as in older models,
however, the interaction is due solely to the LMC, far from the gravitational influence of the
Galaxy (Besla et al. 2010).
A more recent proper-motion study 
of the Clouds encompassing a 450 deg$^2$ area was presented by Vieira et al. (2010, hereafter V10)
based on Southern Proper Motion material (e.g., Girard et al. 2011). V10 were able to
determine the most precise {\it relative} motion between the Clouds due to 
the global solution performed on the contiguous area encompassing the Clouds; this
will help better constrain the recent interaction between the Clouds.
While the size of the proper 
motions for both Clouds are slightly smaller than those from K06 (i.e., less energetic orbits)
the results agree within formal uncertainties. 
See also the Costa et al. (2009) ground-based CCD measurements,
which advocate smaller absolute value proper motions and the discussion about the accuracy of such measurements
in Diaz \& Bekki (2012, hereafter DB12).

Thus, the new picture that the Clouds are on a more energetic orbit than earlier work had
implied, appears real. However, there is still room for debate regarding the first-passage 
scenario proposed by Besla et al. (2007) or a two MW-pericentric passage scenario proposed
by DB12 who explored a large orbital parameter space centered on the
V10 proper motions and realistic MW halo potentials. The consensus is that the MS, LA and Bridge
are of tidal origin, and formed from material pulled out from the SMC in its interaction with 
the LMC and perhaps the MW.
The Galaxy's gravitational influence can help explain the bifurcation (Putman et al. 2003) of the MS, 
as proposed by DB12, and also suggested by the earlier work of Connors et al. (2006). 
In the Besla et al. (2010) model, no such bifurcation is seen.
Moreover, the Besla et al. (2010) model implies that the Clouds will have a close 
encounter in the near future rather than in the near past, as indicated by many other observations.

The main criticism for tidal models is the absence of older stars in the MS and Bridge.
Only young, blue stars were known to exist in the Bridge from the early work of
Irwin et al. (1990), with the inference being that they were born in situ.
Attempts to find older stars in the Bridge and MS as tidally stripped material have failed. 
However, there are recent indications that older stars in the Bridge do exist (Monelli et al. 2011),
and perhaps previous searches of these stars were not wide enough and not necessarily in the right place
(DB12). Another piece of evidence that tidal models are indeed on a sure footing is the 
study of Olsen et al. (2011, hereafter O11) who isolated a population of red giants, supergiants and 
Carbon stars in the LMC with peculiar kinematics. Their radial-velocity analysis indicated that
these stars are either counter-rotating in a plane nearly coplanar with the LMC's disk, or
have  prograde rotation in a disk inclined by $\sim 54\arcdeg$ with respect to the LMC's disk.
They also determined the abundances of 30 such stars as well as 
bona fide LMC stars. They found that the kinematically distinct stars have lower metallicities than do LMC stars;
specifically, [Fe/H]=$-1.25\pm0.13$ dex,
while LMC stars have [Fe/H]=$-0.56\pm0.02$ dex. The kinematical and chemical properties of these stars 
lead O11 to suggest that they were captured from the SMC.

In this study, we focus on two observational aspects that will help better model and understand
the LMC/SMC/MW interaction.
The first is to provide a reliable map of young, OB-type candidate stars over a $\sim 7900$ deg$^{2}$ area
that includes the Clouds' periphery, the Bridge, the LA and part of the MS. We select these 
candidates using photometry from the Galaxy Evolution Explorer survey (GALEX), the 
Two Micron All Sky Survey (2MASS), the Southern Proper Motion Program 4 (SPM4), and the American Association of
Variable Star Observers All Sky Photometric Survey (APASS) and proper motions from SPM4.
The second aspect, is a proper-motion analysis of the kinematically distinct O11 giants, in the LMC.
In this study we have focused on OB-type candidates and giants in order to be able to reliably eliminate
foreground contamination, and to have distant stars with apparent magnitudes in the range where
SPM4 proper-motion errors are well under control.

The outline of the paper is as follows: in the next Section we present the catalogs used, followed by
the selection of OB candidates in Section 3. In Section 4, we present the spatial distribution 
of the candidates in an area focused on the Clouds, then on a larger-scale area encompassing the MS and LA.
In Section 5, we analyze the proper motions of the O11 sample and show that the proper-motions
also indicate a kinematically distinct group reinforcing the conclusions of O11 and further characterizing
the kinematical nature of this population. In Section 6 we discuss our findings in the context 
of the dynamical interaction of the Clouds and we summarize our results in Section 7.

\section{Catalogs}

\subsection{The Southern Proper-Motion Catalog}

The fourth installment of the Yale/San Juan Southern Proper Motion Catalog, SPM4
contains positions, absolute proper motions and $B$, $V$ photometry for over 103 
million objects south of $\delta = -20\arcdeg$. The catalog 
construction is described in Girard et al. (2011); it is roughly complete to $V = 17.5$,
and the proper-motion precision is 2-3 mas~yr$^{-1}$ for well-measured stars.
At the bright end, proper motions are on the International Celestial Reference System
via $Hipparcos$ stars, while the faint end is tied to 
the inertial system via galaxies. 
About two thirds of the catalog include $V$ CCD photometry obtained with 
a 4K $\times$ 4K PixelVision (PV) camera ($0.83\arcsec$/pix) mounted 
in the focal plane of
the yellow lens of the 50-cm double astrograph at El Leoncito, Argentina.
Initially, an Apogee 1 K $\times$ 1 K camera ($1.3\arcsec$/pix) was placed behind 
the blue lens of the astrograph; this was later replaced by a 2 K $\times$ 2 K 
Apogee Alta camera ($0.74\arcsec$/pix). The PV data were used for astrometry and 
$V$ photometry. The data from the Apogee cameras 
were used to obtain $B$ photometry only. Since the $B$ CCD area coverage is 
only about $20\%$ of the PV area coverage, we did not use $B$ photometry in 
the current study. The CCD data were calibrated using $Tycho2$ $BV$ values 
corrected to the Johnson system. While SPM4 also includes photographic photometry, 
we chose not to use it here, as it is less reliable than the CCD data.
Hereafter, when mentioning SPM4 $V$ data, we refer solely to the calibrated PV CCD data.
Internal error estimates (Girard et al. 2011) indicate a $V$ precision
between 0.02 and 0.04 mags. 

To further assess errors in the SPM4 $V$ photometry from an external comparison, 
we make use of the OGLE3 $V$ photometry in the
 LMC\footnote{ftp.astrouw.edu.pl/ogle/ogle3/maps/lmc} 
(Udalski et al. 2008). Since 
the SPM survey uses a diffraction grating in front of the objective, and the area 
under consideration is dense, crowding will affect the SPM4 photometry. Therefore
this comparison will not necessarily reflect the average photometric errors in SPM4,
but rather an upper limit. A total of 111735 objects were matched
between the two catalogs, within 
a $2\arcsec$ matching radius, and with $12 \le V_{OGLE3} \le 18$.
In Figure 1 we show the area coverage of
OGLE3 and that of the intersection of OGLE3 and SPM4 $V$. Isodensity contours of M giants
selected from 2MASS (see Section 2.4) are also shown. The coordinates are 
a gnomonic projection of Galactic longitude and latitude centered
at $(l, b) = (290\arcdeg, -40\arcdeg)$. Throughout this paper, the spatial 
distribution will be represented with these coordinates, unless otherwise
noted.
In Figure 2 we show $V$-magnitude differences as a function of $V_{OGLE3}$,
for every tenth data point. The red line shows a moving median of the differences.
A vertical line at $V_{OGLE3}  = 16.7$ indicates the limiting magnitude of
our OB-star candidates. This shallow limit compared to SPM4's $V=17.5$ limit is
due to the fact that in its construction, SPM4 used several input catalogs
including 2MASS for the faint end (Girard et al. 2011). 
Thus, the shallow $V$ limit in SPM4 reflects the 2MASS limit for these blue objects,
rather than the photographic limiting magnitude of the SPM plates.
From Fig. 2, the scatter in the $V$-magnitude differences is
 $\sim 0.07$ mag for $13 \le V \le 16$ and $0.09$ mag for $ 16 \le V \le 17$. 
Since the dominant error in the
differences is from SPM4 $V$ , we will adopt these values as upper limits of
the photometric errors in SPM4.
The difference between SPM4 and OGLE3 as given by the moving median 
in $13.0 \le V_{OGLE3} \le 16.5$ is $-0.025\pm0.001$ mag.

\begin{figure}
\includegraphics[scale=0.5]{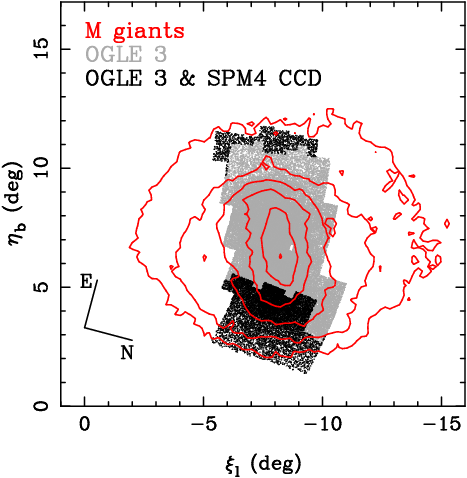}
\caption{Spatial distribution of OGLE3 (grey) and the intersection of OGLE3 and SPM4 (black) in the
LMC. M-giant isodensity contours are shown in red. Coordinates are a 
gnomonic projection of Galactic longitude and latitude centered at $(l, b) = (290\arcdeg, -40\arcdeg)$.}
\end{figure}

\begin{figure}
\includegraphics[scale=0.4,angle=-90]{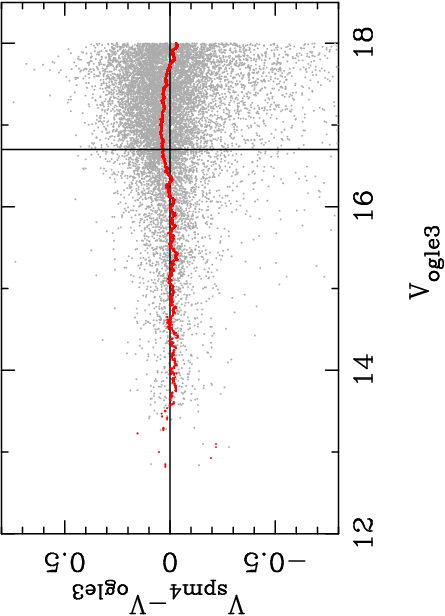}
\caption{$V$-magnitude differences between SPM4 and OGLE3 as a function of
$V_{OGLE3}$. A moving median is shown in red; the vertical line indicates the
limiting magnitude of our OB candidates.}
\end{figure}

Finally, we will also consider here
a subcatalog built only in the area of the Clouds and described in V10.
This catalog uses the same material as SPM4, but
slightly different processing, with more careful treatment of
magnitude-dependent systematics. The V10 catalog has $\sim 1.4$ million objects
 encompassing a 450-deg$^{2}$ contiguous area including the Clouds and the
Bridge (excluding the central regions of the Clouds due to crowding). It was built 
for the purpose of measuring the individual absolute proper motions of the Clouds 
and a very precise relative proper motion between the Clouds. We will also use the V10
proper motions in the selection of OB candidates to compare with the selection 
provided by SPM4 proper motions.

\subsection{The AAVSO All-Sky Photometric Survey}

To increase the area coverage of our study, we supplement the SPM4 $V$ data
with $V$ photometry from the AAVSO All-Sky Photometric Survey - 
APASS\footnote{www.aavso.org/aavso-photometric-all-sky-survey-data-release-1}, 
described in Henden et al. (2011).
Here we use the third data release, which includes Johnson $B,V$, 
Sloan $g,r,i$ and positions for 18.9 million objects. Magnitudes are 
reliable between $V \sim 10$ to 17.0; the pixel scale is 2.57$\arcsec$ and the
field of view is $2.9\arcdeg\times2.9\arcdeg$. The catalog starts to be incomplete 
at $V \sim 16.0$; it is not yet fully contiguous, missing some $10\% - 20\%$ in area.  

Objects were matched by position, using a $2\arcsec$ matching radius. 
We compare $V$ magnitudes from SPM4 with those in APASS, in overlapping regions.
For well-measured stars,
magnitude differences have a scatter of $\sim 0.07-0.08$ mag; assuming errors are
similar in both catalogs, this indicates $V$ errors of 0.05-0.06 mag.
However, there are regions with significant offsets of the order of 0.5 to 1.0 mag;
these regions have the size of the APASS field of view, therefore
we believe they are some unreliably calibrated fields in APASS. We have
carefully examined magnitude differences as a function of Declination and RA,
and we find that large offsets are not present in the overall Clouds and Bridge 
areas; the largest offsets are present at $\delta < -80\arcdeg$.
We therefore proceed to use APASS $V$ data for SPM4 objects that do not have
CCD $V$ data. 
Of the entire list of objects to be examined
$20\%$ have $V$ from APASS; and of the list of OB candidates, $10\%$ 
are from APASS.

\subsection{The GALEX UV Catalog}

Here, we use the GALEX fifth data release (GR5) catalog 
described by Bianchi et al. (2011).
We focus on the all-sky imaging survey (AIS) that provides a list of 63.5 million unique 
sources with photometry in the far UV band 
($FUV,~1344-1786${\AA}, $\lambda_{eff}=1538.6${\AA})
and the near UV band $(NUV,~1771-2831${\AA}, 
$\lambda_{eff}=2315.7${\AA})\footnote{archive.stsci.edu/prepds/bianchi-gr5xdr7/catalogUV.html}.
Only sources with a distance $\le 0.5\arcdeg$ from the field center and
with errors $\epsilon_{NUV} < 0.5$ mag were 
included in this catalog.

We match the GALEX sources with SPM4 objects 
by position using a matching radius of $3\arcsec$. Duplicate matches within this radius
are eliminated by keeping the match with the smallest separation.
In our analysis, we use only the $NUV$ band, since the $FUV$ band has far fewer
detections.

\subsection{2MASS and 6x2MASS}

The SPM4 catalog had been cross-matched with 2MASS (Skrutskie et al. 2006)
prior to release (Girard et al. 2011); therefore $J,H,K$ magnitudes 
are present in the SPM4. In our analysis, we chose to use the $J$ band
as it is the deepest IR band. Since our OB candidates have extremely 
blue colors and relatively faint $V$ magnitudes (see Section 3),
the $J$ magnitudes are at the faint end of 2MASS, and thus prone to
larger errors than the average of 2MASS. To verify
that our OB candidate selection is not seriously affected by errors 
at the faint end of 2MASS, we also make use of the deeper survey 6X 2MASS
in a few selected areas\footnote{www.ipac.caltech.edu/2mass/releases/allsky/doc/}. 
Specifically, we use the 383 deg$^{2}$ and 127 deg$^{2}$ areas in the LMC and SMC respectively.
Thus, in addition to the 2MASS-based selection we present 
a 6X 2MASS-based selection for comparison.

For illustration purposes we construct isodensity contours of the M giants
in the LMC  and SMC; these represent an intermediately old population, of
[Fe/H] $\sim -0.5$.
M giants are selected from 2MASS within a 400-deg$^{2}$ box 
centered on the LMC, and a  225-deg$^{2}$ box centered on the SMC.
M giants are selected based on 2MASS colors according to the 
prescription described in Majewski et al. (2003).

\section{Selection of OB-type Candidates}

We compile a list of about 4 million objects that have $NUV,V,J$ photometry and 
proper motions.
For each object in this list we determine the reddening E$_{B-V}$ from the Schlegel et al. (1998) maps 
and keep for  further analysis objects with $E_{B-V} < 0.5$. 
Objects are dereddened using the relationships from Majewski et al. (2003)
for IR photometry, and those from Kinman et al. (2007) for the $NUV$ band.
Specifically, A$_{NUV}$ = 8.90E$_{B-V}$. For the $V$ band, we used 
A$_{V}$ = 3.1.
The Clouds are known to be affected by differential 
reddening due to the spatial variations of their intrinsic extinction as show by 
e.g., Zaritsky et al. (2004), Haschke et al. (2011). Our study in unlikely to be affected 
by this, as we sample regions in the outskirts of the Clouds where the extinction as well 
as its spatial variation is small according to Haschke et al. (2011).

In Figure 3 (top) we show the $(NUV-V)_0$ versus $(V-J)_0$ diagram, and the 
color-magnitude diagram (bottom).
Grey symbols show our compiled list. The black curve shows model main-sequence colors, while the
green curves are white-dwarf models taken from  Vennes et al. (2010). The white-dwarf 
models correspond to two limiting masses of 0.4 M$_{\odot}$ and 1.2 M$_{\odot}$.
We have also matched our list with catalogs of known-type objects, to check their
color distribution. These are the QSO and AGN catalog by  V\'{e}ron-Cetty \& V\'{e}ron (2006)
(dark blue), RR Lyrae stars from the All Sky Automated Survey (ASAS) Szczygiel et al. (2009) (light blue),
white dwarf candidates selected photometrically from GALEX and SDSS (Binachi et al. 2011) (green),
and OB stars spectroscopically identified from Rolleston et al. (1999), Reed (2003,2005), 
Bonanos et al. (2009, 2010) (in red). 
The initial color selection is shown as the black box with 
$-2.0 \le (NUV-V)_0 \le 2.0$ and $ -1.2 \le (V-J)_0 \le -0.2$.
We also trim by magnitude errors: $\epsilon_{NUV} \le 0.10$ mag and
$\epsilon_J \le 0.15$ mag. This selection implies that the latest stellar type is B8 according
to the models of Vennes et al. (2010). The expected contamination in this color range is due to foreground 
stars such as OB subdwarfs and white dwarfs. Our selection provides 3099 objects of which only 
two are classified as non-stellar objects: one is a 2MASS extended source, the other a LEDA galaxy.

In Figure 4 we show the spatial distribution in Galactic coordinates of this sample (top),
and the $V_0$ distribution as a function of Galactic $l$ (middle) and $b$ (bottom).
Between  $l = -90\arcdeg$ to $-55\arcdeg$ and $b = -50\arcdeg$ to $-25\arcdeg$ one can see
as a density enhancement the periphery of the LMC and SMC's wing. 
Density enhancements also appear at low latitudes, most likely representing disk populations.
This is also apparent in the middle and bottom panels of Fig. 4; the Clouds' population 
sets in at $V_0 \sim 13$. 
Therefore we keep only objects with $V_{0} \ge 13.0$. 
The faint limit of our color-selected sample is $V_0 \sim 16.5$. Incompleteness begins in 
at $V_0 \sim 15.5$, primarily due to 2MASS.

\begin{figure}
\includegraphics[scale=0.48]{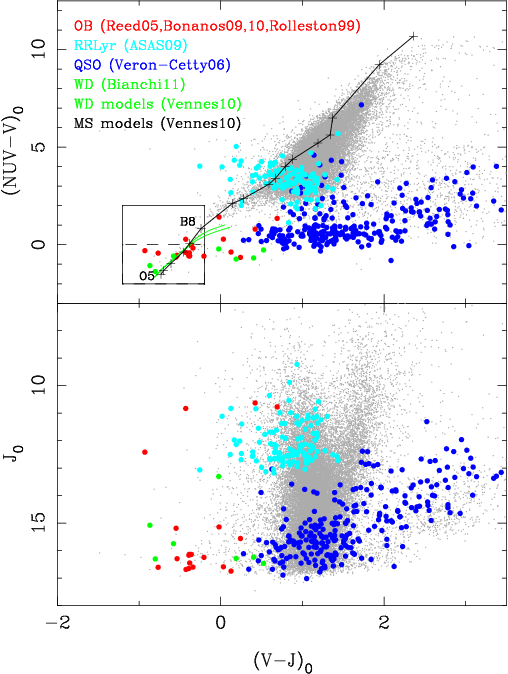}
\caption{Color-color (top) and color-magnitude (bottom) plots of our list of objects with
UV, optical and IR photometry and with proper motions (grey symbols).
Main-sequence  models are shown with a black line, and white-dwarf models with a green line.
Objects of known type from various catalogs (see text) matched with our list are also
shown and labeled. The preliminary color selection is represented by the black box; the more restrictive final
color selection is represented with a dashed line.}
\end{figure} 
\begin{figure}
\includegraphics[scale=0.45]{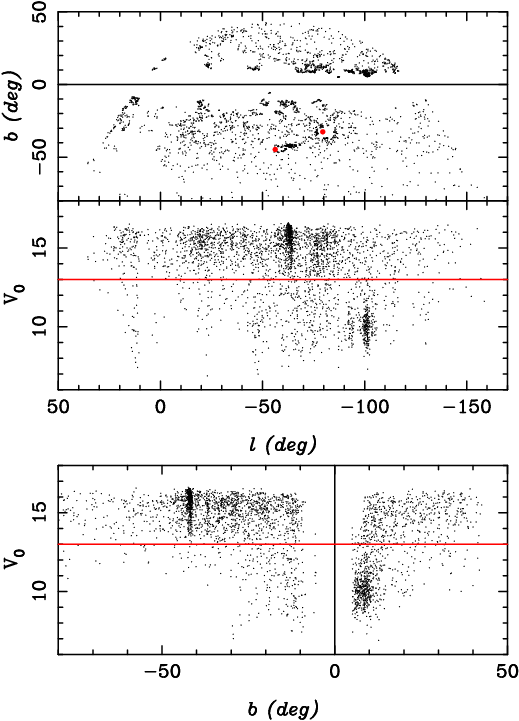}
\caption{Spatial distribution of the color-selected sample
$-2.0 \le (NUV-V)_0 \le 2.0$ and $ -1.2 \le (V-J)_0 \le -0.2$ (top), and
$V$-magnitude distribution as a function of Galactic coordinates. The red line
indicates the $V_0$ magnitude cut intended to eliminate nearby, foreground objects.
The centers of the Clouds are marked with a red dot in the top panel.}
\end{figure}
\begin{figure*}
\includegraphics[scale=0.75,angle=-90]{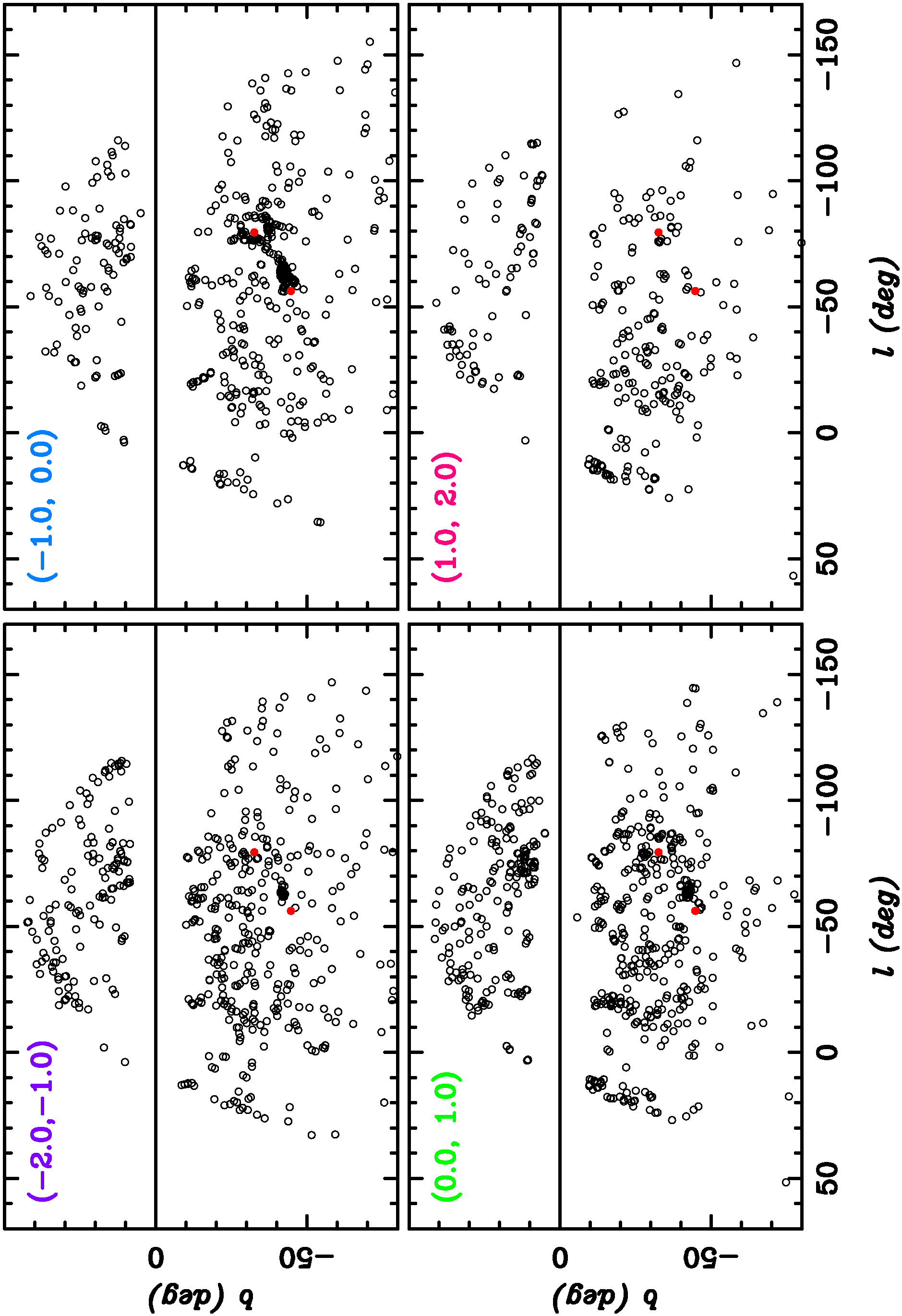}
\caption{Galactic-coordinate distributions of four 
subsamples selected by 
$(NUV-V)_0$ color. The overall sample
comes from the $ -1.2 \le (V-J)_0 \le -0.2$, 
$V_0 \ge 13.0$ mag cuts. The $(NUV-V)_0$ interval is specified in each panel.
The centers of the Clouds are marked with a red dot.}
\end{figure*}
\begin{figure}
\includegraphics[scale=0.42,angle=-90]{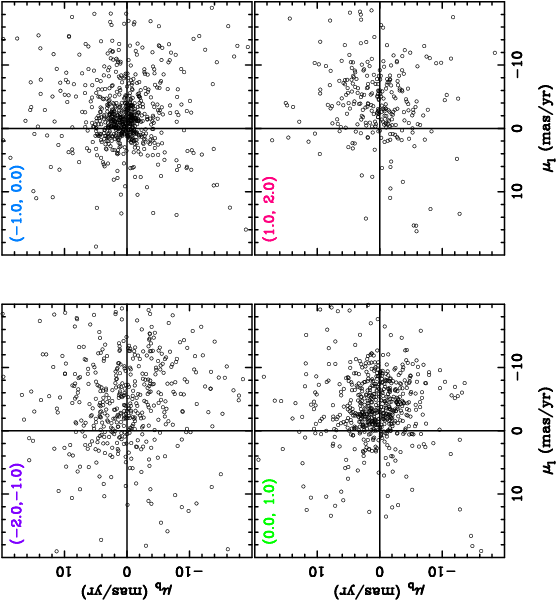}
\caption{Proper-motion distributions of the four $(NUV-V)_0$ subsamples shown in Fig. 5.}
\end{figure}
\begin{figure}
\includegraphics[scale=0.37,angle=-90]{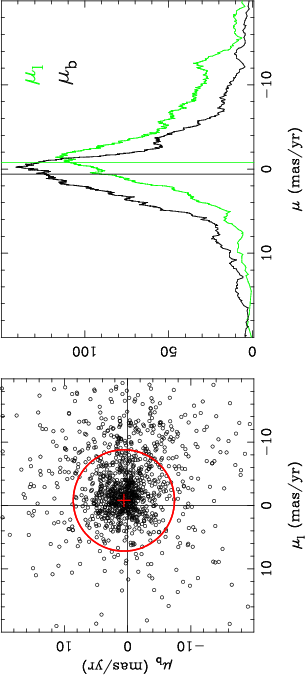}
\caption{Proper-motion selection for the $-2.0 \le (NUV-V)_0 \le 0.0$, $ -1.2 \le (V-J)_0 \le -0.2$
and $V_0 \ge 13.0$ mag sample (left). 
The red circle shows the trimming radius of 8 mas~yr$^{-1}$ centered on the average value (red cross) for 
stars located in the SMC wing.
The proper-motion distributions in each coordinate are
shown in the right panel, together with the central value (vertical lines) used in the proper-motion cut.}
\end{figure}

Next, we divide our color-selected sample into four $(NUV-V)_0$ subsamples and explore their 
spatial and proper-motion distributions. From Figure 5, the most prominent Clouds sample
is in the color range $(-1.0 < (NUV-V)_0 < 0.0)$, while the reddest subsample appears 
devoid of Clouds stars. From the proper-motion distributions shown in Figure 6, 
it is also clear that the color range $(-1.0 < (NUV-V)_0 < 0.0)$ has a tight 
proper-motion clump centered around $(\mu_l, \mu_b) \sim (-1.0,1.0)$ mas~yr$^{-1}$ 
and is indicative of a kinematically cold and
distant population. We therefore choose a stringently-defined 
sample by restricting the color range $(-2.0 < (NUV-V)_0 < 0.0)$, i.e., selecting the bluest two
susbsamples shown in Figs. 5 and 6. This will limit our sample to spectral types
earlier than B5 ($T_{eff} > 16,000$ K). By limiting to the bluest colors, we aim to target only the 
youngest distant stars, as well as to avoid some of the foreground contamination
due to OB subdwarfs, which are known to be less frequent at bluer colors
(e.g., Vennes et al. 2010).
To eliminate nearby stars (i.e. white dwarfs) we also 
trim in proper-motion space as follows.
We determine the average proper motion of stars located in the SMC wing, then we
keep only stars within a proper-motion radius $ \le 8.0$ mas~yr$^{-1}$ from this mean
and with proper motion errors $\epsilon_{\mu_l} \le 4.0$ and  $\epsilon_{\mu_b} \le 4.0$ mas~yr$^{-1}$.
Figure 7 shows the proper-motion cut for the color and magnitude-selected sample (left), and the
proper-motion distribution in each coordinate (right).
There are 567 stars in this final sample, which was constructed to be as free as possible from 
foreground contamination. Some OB-type stars that are genuine members of the
Clouds's system will be missed due to the
circular footprint of GALEX, as well as due to the lack of photometry (mostly UV, but some $V$ as well,
see next Section).

\section{Spatial Distribution of OB-type Candidates}

\subsection{The Clouds and Bridge Region} 

In Figure 8 we show the gnomonic projection of the
spatial distribution of our OB-type candidates,
in a 2500-deg$^2$ region centered at $(l,b) = (290\arcdeg, -40\arcdeg)$.
Data combinations from various surveys are also highlighted in different colors as follows:
GALEX and $V$ from SPM4 in green, GALEX and $V$ from APASS in yellow, GALEX and no $V$
data in orange. White gaps are due to the lack of GALEX data: note the round footprint
of the GALEX catalog. 2MASS data and SPM4 proper motions are practically everywhere in this area,
and thus not highlighted. The location of the Clouds is represented with 
M-giant isodensity contours from 2MASS (see also Section 2.4). Our OB-type candidates are shown with 
blue circles.
We have also applied the same selection criteria, but replacing the $J$-magnitude from 2MASS,
with that from 6X 2MASS, for $J > 13.0$ mag. The resulting sample contains 586 stars, i.e.
19 more. The spatial distribution is essentially unchanged from that of 
the 2MASS-only sample. The 6X 2MASS sample is shown in Figure 9; 
the two clumps at the end of the LMC bar appear slightly more enhanced than in the 2MASS sample.
 
Finally, we perform one more test. We replace the SPM4 proper motions with
the proper motions from the catalog of V10 based on the same material as SPM4,
but with slightly different processing. We regard this latter catalog as more accurate and 
precise than the SPM4, since more careful treatment was possible in this 
restricted area. The V10 catalog covers a contiguous 450-deg$^2$ region encompassing the Clouds, 
and excluding the inner regions of the Clouds.
In Figure 10 we show the same data from Fig. 8, only with the OB candidates being selected 
in proper-motion space using the V10 catalog and the same criteria as in the previous Section.

In both cases, the spatial distribution of OB candidates shows a very well-populated SMC wing, with
two branches partly surrounding the SMC; the opposite side from the SMC wing,
i.e., the north-west periphery of the SMC appears devoid of these stars.
From the SMC wing a rather narrow ($\sim 3\arcdeg$) band extends toward the LMC,
ending some $\sim 4\arcdeg$ from the LMC periphery, or at RA $\sim 3.4^{h}$. 
A group of 2-3 candidates 
is present in the periphery of the LMC and in the direction of the band coming
from the SMC wing at $(l,b) \sim (286.3\arcdeg,-36.9\arcdeg)$,
(RA, Dec) = $(4.2^{h},-72.9\arcdeg)$; perhaps this indicates
the spot where material from the
SMC is funneled into the LMC after the close encounter between the Clouds. 
While there appears to be a gap of $\sim 4\arcdeg$
from the easternmost end of the SMC wing to the group of 2-3 stars in the LMC's
periphery, it is also possible that we miss candidates due to the GALEX gaps
(see Fig. 8).

The LMC periphery is well populated with these OB candidates, with some of them 
scattered at nearly $10\arcdeg$ from the LMC's center. Most remarkably,
there are two clumps of these stars at each end of the bar, some $5\arcdeg$ from the
LMC's center. In the north-west clump there are 19 stars (23 if we use $6\times$ 2MASS),
and in the south-east clump there are 8 stars (9 if we use $6\times$ 2MASS).
The V10 proper-motion catalog excluded inner regions of the LMC, 
thus, these areas in the LMC are not well represented in V10.
Also, a clump is seen in an arm-like feature, where known young associations 
and clusters exist (see below). The northern side of the LMC (which roughly
corresponds to the near side)  appears to have a higher 
density of these stars than the southern side.

\begin{figure}
\includegraphics[scale=0.52]{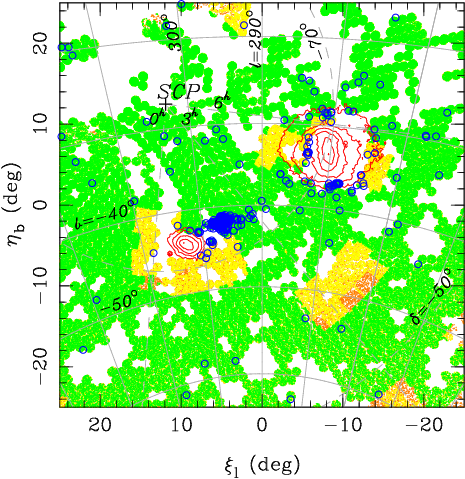}
\caption{Spatial distribution of OB-type candidates (blue circles).
Regions with GALEX and $V$ from SPM4 are shown in green, regions with GALEX and $V$ from APASS in 
yellow, and regions with only GALEX photometry are shown in orange. M-giant isodensity contours in the Clouds'
areas are shown in red.}
\end{figure}

\begin{figure}
\includegraphics[scale=0.52]{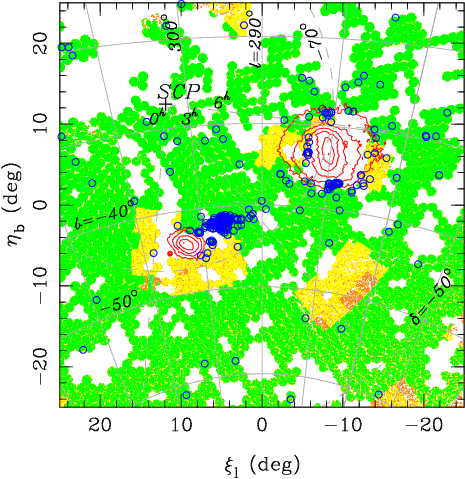}
\caption{Same as Fig.8, except that the 2MASS data have been replaced with 6X 2MASS data
for $J > 13.0$ mag in making the selection of OB-type candidates.}
\end{figure}

\begin{figure}
\includegraphics[scale=0.52]{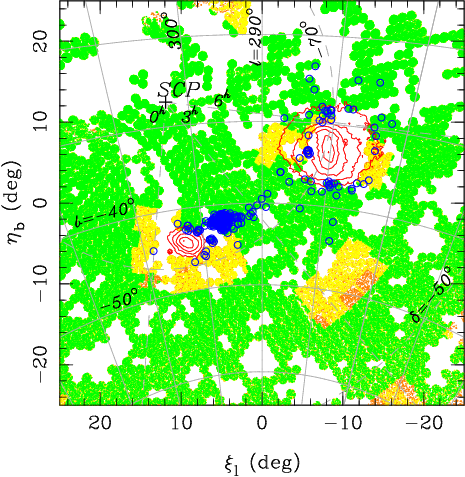}
\caption{Same as Fig.8, except that the SPM4 proper motions have been replaced with
the proper-motion catalog of V10, which covers a 450-deg$^2$ contiguous area
encompassing the Clouds.}
\end{figure}

\begin{figure}
\includegraphics[scale=0.52]{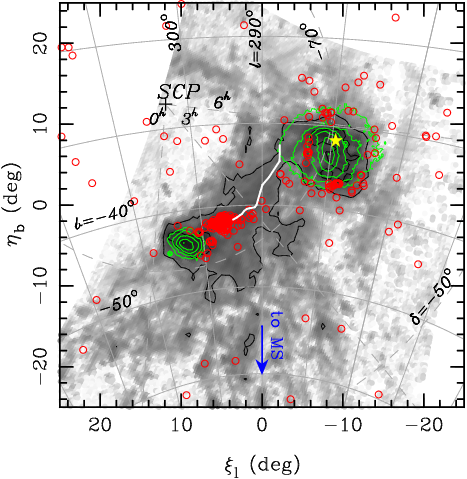}
\caption{OB candidates (red circles) against the H I map from Putman et al. (2003) - grey scale. The black contours
correspond to $N_{HI} = 10^{20}$
 and $10^{21}$ atoms~cm$^{-2}$; the latter one being barely visible in the
inner regions of the LMC. M-giant isodensity contours are
shown in green. 
The highest density region in the Bridge is highlighted with a ridge line
shown in white. The 30 Doradus star-forming region is shown with a yellow star symbol.}
\end{figure}

In Figure 11, we show the H I map from Putman et al (2003) in the same region
depicted in Figs 8-10, together with our OB candidates from Fig. 8 
(i.e., SPM4 proper motions, and 2MASS photometry). 
Throughout the remainder of this paper, we will use this sample.
Notably, the band of OB candidates extending from the SMC wing toward the LMC
(i.e. at RA $> 3^h$) is offset by $ 1\arcdeg - 2\arcdeg$ from 
the highest $N_{HI}$ density region in the bridge shown with a white line 
in Fig. 11. This also holds for
the group of three candidates in the periphery of the LMC, 
at $(l,b) \sim (286.3\arcdeg,-36.9\arcdeg)$, (RA, Dec) = $(4.2^{h},-72.9\arcdeg)$.

Next, we compare our list of 567 candidates with  the Bica et al. (2008) catalogs of
emission nebulae (1445 objects), stellar associations (3326) and clusters
(3740) in the Clouds and Bridge. We search for positional coincidence among our
OB candidate list and each of the Bica et al. (2008) catalogs; matched objects
have separations that are smaller than or equal to the average radius of
the extended object (i.e., 0.5$\times$(semimajor+semiminor axis) from Bica et al. 2008).
Clusters in this catalog are those older than 5 Myr.
We find 15 matches with the emission nebulae catalog, with 3 in the LMC, and the rest in the SMC wing;
71 matches with the associations catalog, with 2 in the SMC, 11 in the LMC and the rest in the Bridge and SMC wing;
and 10 matches with the cluster catalog, with 2 in the LMC, 1 near the edge of the SMC and the rest 
in the SMC wing. The distributions of each of these catalogs together with our OB candidates are
shown in Figs 12, 13 and 14. Most of the matches are in the SMC wing/Bridge, and in the
clump of stars to the left of the LMC bar, where clusters also show an enhancement
resembling a spiral arm. The two clumps found at the ends of the LMC bar appear farther out
than the extent of the emission nebulae and associations. In fact, none of the matches with all
three catalogs are in these two clumps at the ends of the bar.
In the SMC region, our candidates that form the two branches that partly 
surround the SMC do not seem to be
the edges of a roundish distribution of associations/clusters within the SMC.
Emission nebulae, and associations and even clusters in the SMC have an elongated distribution 
along the SMC bar, which is oriented nearly perpendicular to the direction to the SMC wing.
The configuration seen in our OB candidates around the SMC is also apparent in the 
study of Gardiner \& Hatzidimitriou (1992) where blue stars were selected from the UKST 
photographic survey. Their Figure 5 indicates that the periphery 
(i.e., out to $\sim 2\arcdeg$ of the SMC center)  is partly
surrounded by blue stars; our study shows the same configuration is present out to radii
of $3\arcdeg - 4\arcdeg$.

\begin{figure}
\includegraphics[scale=0.48]{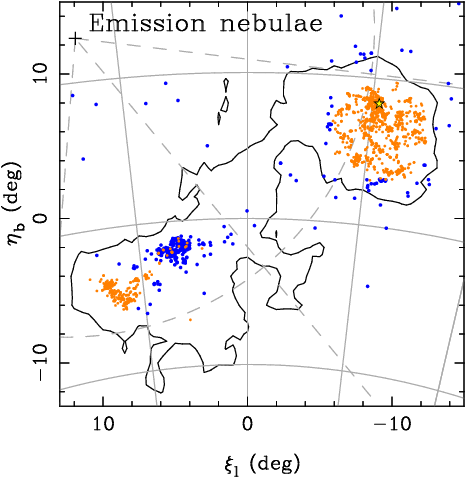}
\caption{Distribution of emission nebulae from Bica et al. (2008) (orange), and our 
OB candidates (blue). The black contour represents the HI isodensity 
of $N_{HI} = 10^{20}$ atoms~cm$^{-2}$. 30 Doradus is shown with a yellow star symbol.}
\end{figure}
\begin{figure}
\includegraphics[scale=0.48]{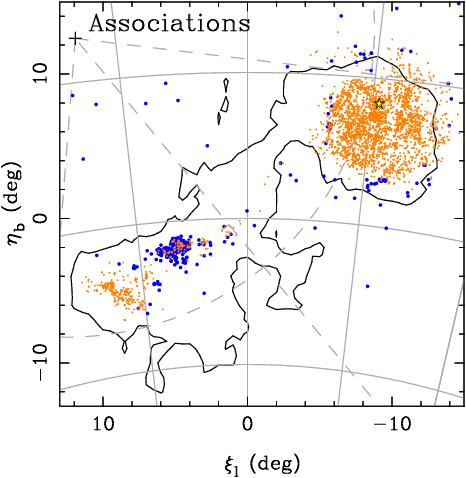}
\caption{Same as in Fig. 12, but with stellar associations shown in orange.}
\end{figure}
\begin{figure}
\includegraphics[scale=0.48]{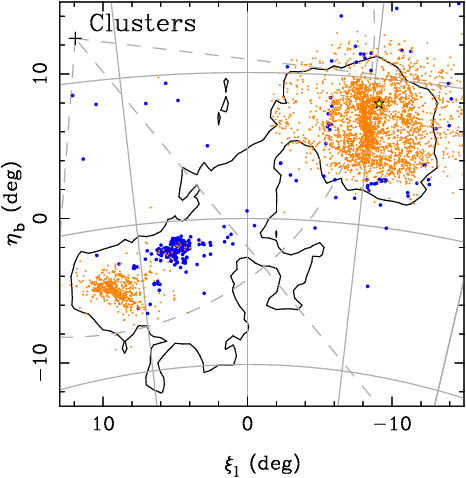}
\caption{Same as in Fig. 12, but with clusters shown in orange.}
\end{figure}

To check possible differences in the population of OB candidates, we separate the Clouds region into  
a) an SMC region which includes the SMC wing
and its extension to RA $\sim 3.4^h$, and the branches surrounding the SMC,
and b) the LMC region, which includes the periphery out to a radius of about
$8\arcdeg$. We calculate the average $V_0$, $(NUV_V)_0$, and $(V-J)_0$ and list these in
Table 1. There is indication that the LMC sample is slightly brighter than the SMC one,
and marginally bluer. The magnitude difference can easily be accounted for by
the different distance moduli of the Clouds, i.e., $\Delta m = 0.49$ if
the distance to the LMC is 50.1 kpc and to the SMC is 62.8 kpc (van der Marel et al. 2002, hereafter
vdM02).
Thus, at a first look, there is no indication that the two groups consist of
significantly different stellar populations.

\begin{table}[htb]
\caption{Average magnitudes and colors}
\begin{tabular}{lrr}
\tableline
\\
    & \multicolumn{1}{c}{SMC region} & \multicolumn{1}{c}{LMC region} \\
\tableline
 $N$          & 270 &   71 \\
 $<V_0>$      & $15.51\pm0.04$   &  $15.08\pm0.09$ \\
 $<(NUV-V)_0>$ & $-0.62\pm0.02$  &  $-0.70\pm0.04$ \\
 $<(V-J)_0>$   & $-0.48\pm0.01$  &  $-0.58\pm0.02$ \\
\tableline
\end{tabular}
\end{table}

\subsection{The Magellanic Stream and Leading Arm} 

In Figure 15 we show the spatial distribution of our OB candidates in MS coordinates
(e.g., Nidever et al. 2010). These are overplotted onto
the H I column density map from Nidever et al. (2010) in the top panel, and 
onto our survey coverage in the bottom panel.
There appear to be $\sim 10$ scattered candidates along the MS;
only a few of these are in high $N_{HI}$ regions. 
There are 4-5 candidates in the region opposite the SMC wing;
these extend from  $\Lambda_{MS} = -40\arcdeg$ to $-15\arcdeg$,
and $B_{MS} = -15\arcdeg$ to $-28\arcdeg$.
More prominent clumps of OB candidates can be found in the
LA regions at $(\Lambda_{MS},B_{MS}) = (16\arcdeg, -20\arcdeg)$,
below the Galactic plane, and at $(\Lambda_{MS},B_{MS}) = 
(\sim 40\arcdeg, -10\arcdeg$ to $0\arcdeg)$, above the Galactic plane.
These clumps coincide with regions of moderate column density. 
Note however the gap in the survey for the clump
below the Galactic plane: specifically the high $N_{HI}$ region at
$(\Lambda_{MS},B_{MS}) = (15\arcdeg, -25\arcdeg)$ is not covered by our data.
Since these clumps are at low Galactic latitude, we may
expect foreground contamination from stars in the MW disk.
The clump below the plane is at 
$(l,b) = (298.2\arcdeg,-12.2\arcdeg)$ and contains $\sim 10$ stars, while the 
grouping above the plane is
at $(l,b) = (276.1\arcdeg-292.5\arcdeg, 11.7\arcdeg)$, and contains some 16 stars.

The Galactic longitudes of these two groups encompass the Carina-Sagittarius arm
 and possibly the Perseus arm at large distances from the Sun
(Vall\'{e}e 2005, Carraro \& Costa 2009, Baume et al. 2009). 
Carraro \& Costa (2009) study the region at $l = 290\arcdeg$ where they find
three distinct groups of young stars: two of them at 
2.5 and 6.0 kpc they believe belong to the Carina-Sagittarius arm
as the line of sight crosses this arm twice, while the 
third at $\sim 12.7$ kpc is believed to belong to the Perseus arm.
If we use these distances and the Galactic latitude of $|b| \sim 12\arcdeg$
of our two groups, we obtain distances from the Galactic plane of
0.5, 1.2 and 2.6 kpc respectively. Thus, regardless of the stellar-population type, 
these stars, if postulated to be at the location of the spiral arms, appear rather far from the Galactic plane.
One could invoke a warped disk to place material so far from the plane. However 
this explanation is also problematic, because
the two clumps are below {\it and} above the plane at similar Galactic longitudes.

The most plausible contamination of our sample by foreground stars in this particular
direction is from nearby stars (i.e., closer to the plane), 
possibly white dwarfs.
On average, $V_{0} = 14.6$ mag (with a range from $\sim 14$ to 16 mag) for
our OB candidates in these two groups. Assuming $M_V \sim 10.0$ for white dwarfs
(e.g., Vennes et al. 2010 models), we obtain a distance of the order of 100 pc.
To check the white-dwarf frequency, we use the Besancon-model (Robin et al. 2003)
predictions for the below-the-plane location of our clump of OB candidates. 
Specifically, in an area of 6 deg$^{2}$ we have 10 OB candidates.
In the same area, the Besancon model predicts only one white dwarf that satisfies our color and
magnitude criteria. However, its proper motion is $\sim 100$ mas~yr$^{-1}$, and therefore would 
not have made our proper-motion cut.

We conclude that it is not reasonable to explain the density enhancements of OB candidates
at low latitudes as spiral arm constituents, nor as foreground contamination from 
white dwarfs.  
Clearly, spectroscopic followup is necessary to elucidate this further.
Nevertheless, our OB candidates suggest the existence of 
recently formed stars in the LA, an intriguing possibility that has bearing on its formation.

\begin{figure*}
\includegraphics[scale=0.9]{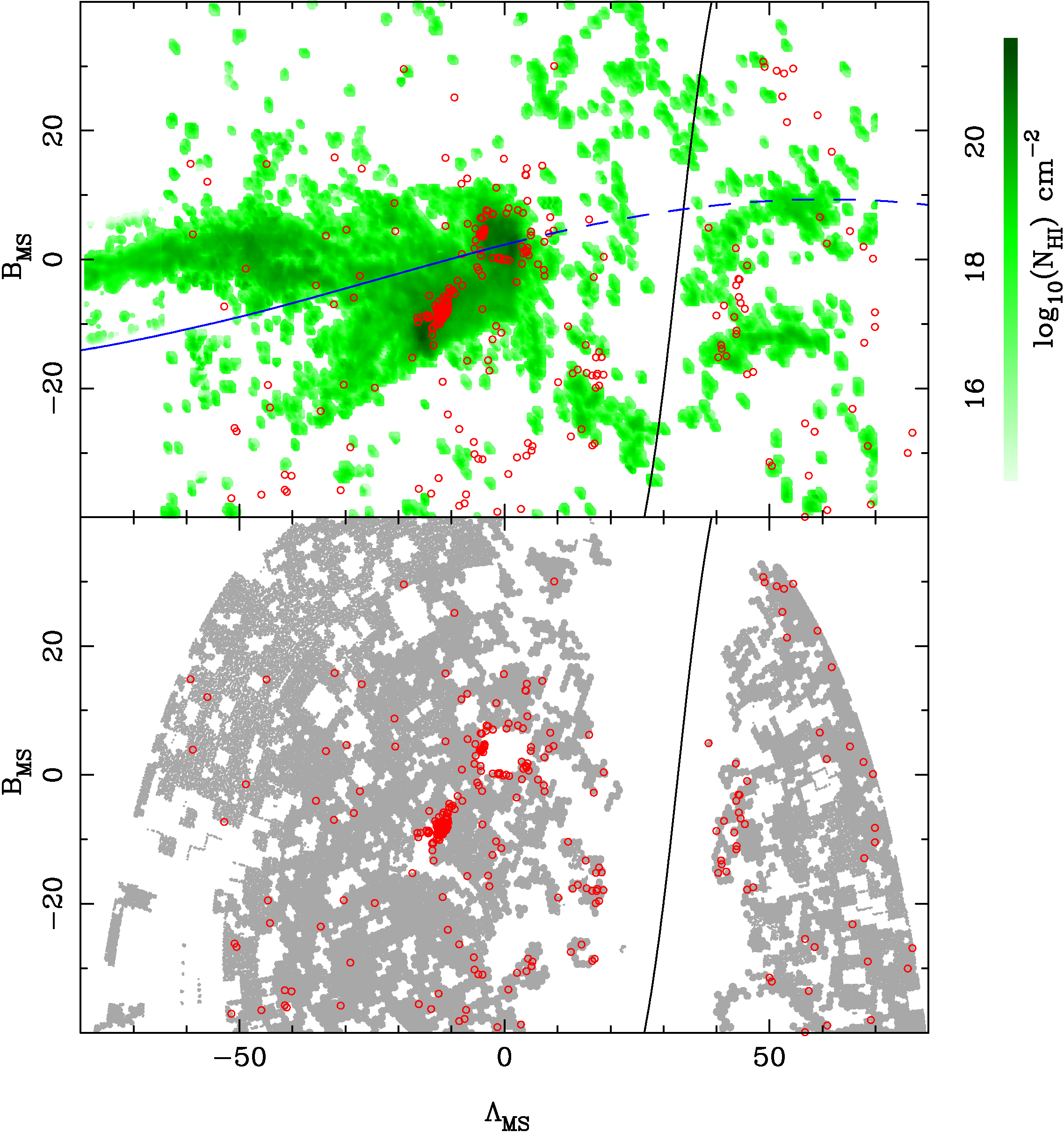}
\caption{Spatial distribution in Magellanic Stream coordinates of our OB candidates (red).
The top panel shows the H I column density distribution from Nidever et al. (2010)
(green) while the bottom shows the area coverage of our study (grey). The Galactic plane 
is shown with a black line. The orbit of the LMC is shown with a dark blue line, backward in time
(continuous) and forward in time (dashed).}
\end{figure*}

\section{Proper Motions of the Olsen et al. (2011) Sample}

\subsection{A Kinematically Distinct Population?}

Recently, O11 presented a radial-velocity study of a 
sample of $\sim 5900$ red supergiants, oxygen- and carbon-rich asymptotic giant branch (AGB) stars and
other giants.
Examining the outliers from the fit of the LMC rotation curve, they find a 
population of stars that have radial velocities that are kinematically distinct.
Based on radial velocities, O11 suggest these stars
can either counter-rotate in a plane closely aligned with the LMC disk, or
rotate in the same sense as the LMC disk, but in a plane inclined by $\sim 54\arcdeg$ to the LMC disk.
This population represents $5\%$ of the total population. O11  measure the 
metallicities of 30 such stars and find that they are 
also distinct, with [Fe/H] = $-1.25\pm0.13$ dex when compared to the bona fide LMC stars which have
[Fe/H] = $-0.56\pm0.02$ dex. Their inference is that these stars were captured from the SMC, specifically
from the periphery of the SMC.

Here, we use the O11 catalog matched with the SPM4 catalog to explore whether a proper-motion 
difference is present in the two samples. The majority of the O11 data lie in the inner regions 
of the LMC where proper-motion measurements are most challenging due to crowding.
We develop a differential approach to assess any proper-motion differences.
We match 285 stars from the O11 ``captured'' sample (hereafter called captured-SMC)  
and 4138 from the ``bona fide LMC'' sample (hereafter called LMC) with SPM4, using a matching
radius of $1.5\arcsec$. In Figure 16 we show the spatial distribution of 
the two samples in the same coordinate system as in Figs 8-10. For reference, the
M-giant isodensity contours are shown in black. The LMC's center 
as adopted from vdM02 is also marked, as well as the line of nodes of the LMC disk with the orientation
adopted from O11 (blue line).

To determine the proper-motion difference between the two samples, one can in principle
take the average values for both samples; in practice, magnitude-dependent systematics that also
vary with spatial location are likely to affect such a direct determination.
For this reason, we develop a local, differential solution as follows. The LMC sample is chosen
as a reference sample, while the captured-SMC as a target sample. For each target
star, we define a local sample formed from $N_l$ neighboring, reference stars, and calculate the
proper-motion difference between target and the average of the reference stars. 
Then, we take the average of these differences for all target stars after trimming outliers. 
We explore ranges for 
two parameters upon which such a solution
might depend: $N_l = 50$, and 200 and the proper-motion box size for trimming outliers in the target sample, i.e.,
$\pm20$ mas~yr$^{-1}$ and $\pm40$ mas~yr$^{-1}$ in both proper-motion coordinates.
We also perform three types of local solutions which differ by the definition of the ``distance''
between reference and target stars. 
The first solution uses a metric that is a combination of spatial distance and $V$ magnitude difference.
The spatial distance between target and reference is given by the gnomonic projection of
celestial coordinates at the location of the target star. 
The metric distance is defined as $(\xi^2+\eta^2+(\Delta~mag\times10.0)^2)^{1/2}$. Here, 
$\xi$ and $\eta$ are the spatial coordinates in degrees, $\Delta mag$ is the V-magnitude difference
between target and reference star, and the factor 10.0 is used to scale magnitude units and spatial units.
For each target star, reference stars are sorted in increasing order of the defined distance,
and we keep either 50 or 200 reference stars, to represent two different ``sizes'' of the 
local reference system.
The second local solution uses a strict spatial distance, and keeps 
the closest 50 or 200 reference objects that also have 
a $\Delta mag \le 1.0$ mag.
The third local solution uses a strict spatial distance; however 
reference stars are also required to have $\Delta mag \le 1.0$ mag and 
$\Delta (J-K) \le 0.3$ mag. The color restriction aims to 
minimize color-dependent systematics. The color-magnitude diagram of 
captured-SMC/targets stars and LMC/reference stars shown by O11, their Fig. 6,
indicates that the LMC sample contains some bluer stars that are not present in the captured-SMC sample;
thus a color-related analysis is a useful check.

With this procedure, we obtain on average (all solutions and parameters) 
$\Delta \mu_{\alpha} \sim -0.8\pm0.6$  mas~yr$^{-1}$ and
$\Delta \mu_{\delta} \sim -0.2\pm0.6$ mas~yr$^{-1}$. 
Unsatisfied with the rather large uncertainties, we decided to include more information to this
determination. We use the formal proper-motion errors from the SPM4 catalog for target stars, and thus
determine an error-weighted average. This procedure however is prone to any residual proper-motion systematics and, thus, 
needs to be checked. To do so, we use the LMC sample as a test sample. We randomly choose 285 stars 
(mimicking the captured-SMC stars) and treat it as a control sample. Since the LMC sample contains 4138
stars, one can devise 14 independent such control samples. The final result
for a default control sample is the average of these 14 independent control samples.
We apply the same procedures described above, and in principle one should 
obtain proper-motion differences equal to zero. This is readily obtained when we use 
simple averages for target control stars; however, when we use an error-weighted average,
we obtain a non-zero value in $\Delta \mu_{\delta}$. We believe this is due to 
unaccounted for systematics in the SPM4. Finally, we use the proper-motion differences 
as obtained by the overall control sample as a zero point offset to be applied to
the results obtained with the captured-SMC sample. All these results are
summarized in Table 2, where we list the type of solution, the proper-motion difference 
for the captured-SMC sample and the difference for 
the control sample which is subtracted from the previous number as a zero-point
correction. Uncertainties are shown in parentheses. The outcome of the zero-point correction and 
its formal uncertainty are also given.
In Figure 17, we plot all these zero-point corrected determinations and their formal uncertainties; they represent the proper-motion
difference between the captured-SMC sample and the bona fide LMC sample, at the location of each 
captured-SMC star.
As our final value we take a simple average of all these 24 determinations, and as uncertainty 
the average uncertainty as given by the weighted-average solutions.
This number is $\Delta \mu_{\alpha} = -0.51\pm0.30$  mas~yr$^{-1}$ and
$\Delta \mu_{\delta} = -0.03\pm0.29$  mas~yr$^{-1}$; it is represented with a red symbol in Fig. 17.
Thus, we obtain a proper-motion difference consistent with zero along declination and $1.7\sigma$
different from zero along right ascension, which, incidentally, is oriented in the direction of the SMC.

\subsection{Kinematical Implications}

Having evidence that the two samples are kinematically distinct, we proceed to analyze 
the kinematics in a physically meaningful picture. From radial velocities alone and assuming
circular motions, O11 proposed two kinematical solutions for the captured-SMC stars:
one is a nearly coplanar counter-rotating disk, the other a prograde disk inclined  $54\arcdeg\pm2\arcdeg$ from the
LMC disk. Here, we test which of these is preferred given the proper-motion information.

We remind the reader that what we have measured are proper-motion differences between
captured-SMC stars and LMC disk stars, locally, i.e., at different regions in the disk
which itself rotates. In addition to the rotation of the LMC disk, the bulk space motion of the LMC  
will be projected onto the proper motions in variable amounts across the disk of the LMC. 
This is also known as the perspective viewing effect on the proper motions.
Other effects include the precession and nutation of the LMC disk; however the amount of precession for 
instance is rather uncertain, with some studies considering it significant (e.g., vdM02),
while others negligible (e.g., K06). The equations for the velocity field across the LMC are given in 
vdM02, and we use these in a simplified version, i.e., in a coordinate system
parallel and perpendicular to the line of nodes. Here, we will neglect the effect of precession/nutation.
Also, the space motion of the LMC contributes to the proper motion perpendicular to the 
line of nodes as a constant (i.e., does not depend on the distance from the center of mass).
However, the space motion of the LMC has an effect on the proper motion along the line of nodes
that slowly varies with the distance from the center of mass.
From Fig. 16, it is apparent that the distribution of captured-SMC stars has two concentrations,
one in the north-west and the other in the south-east, roughly aligned with the 
line of nodes.  For all these reasons, we define a new coordinate 
system, centered on the LMC's center, and rotated from the equatorial system by $52\arcdeg$,
i.e., with $\xi$ coordinate along the line of nodes, and positive toward south-east, and 
$\eta$ coordinate perpendicular to it, and positive toward north-east. We rotate the
proper-motion differences to this system, having now a component parallel
to the line of nodes $\mu_{\parallel}$, and another perpendicular to it $\mu_{\perp}$.
In this system,  we subtract the contribution of the LMC center of mass proper motion, such that
the proper motion at each location reflects only the disk rotation and the perspective viewing; note that
perspective viewing does not affect the run of $\mu_{\perp}$ with $\xi$.
The LMC disk rotates clockwise, with the near side in the north-east.
Along the line of nodes, at negative $\xi$, the rotation is projected onto  $\mu_{\perp}$ with negative
values; the converse is true for positive $\xi$.

We restrict the captured-SMC sample to stars within $| \eta | \le 3\arcdeg$ (i.e., near the line of nodes),
and $1\arcdeg < | \xi | \le 5\arcdeg$ (i.e., eliminate the few sparse objects in the periphery of the LMC,
and a few objects near the center). We also discard stars with proper-motion differences $| \Delta \mu_{\parallel,\perp} | > 30.0$  mas~yr$^{-1}$.
In principle, our  method is applicable, if all stars are distributed along the line of nodes;
however we have a box aligned with the line of nodes, of thickness $6\arcdeg$ and length $10\arcdeg$.
Given that the orientation of the line of nodes ranges by about $20\arcdeg$ from various studies 
(e.g., it is $142\arcdeg$ in 
O11, and $122\arcdeg$ in vdM02), our box of given thickness along the line of nodes 
can be thought of as absorbing this range. 
The objects are distributed roughly equally at positive and
negative  $\eta$ coordinate. Therefore, we believe that our approximation works well in the context of our 
proper-motion errors. These cuts restrict our sample to between $\sim 200$ and 210 captured-SMC 
stars for the various solutions shown in Table 2.

To calculate the proper motions in this new system, we need a model for the rotation of the LMC disk, as well as
the tangential and line-of-sight velocity of the LMC's center of mass, and the angle between the line of nodes
and the direction of the tangential velocity.  
For the rotation, we adopt the model described in O11, in which the disk has an inclination angle of 
$34.7\arcdeg$ (van der Marel \& Cioni 2001),
a rotation curve that increases linearly from 0 to 87 km~s$^{-1}$ between 0 and 2.4
kpc from the LMC's center and is flat afterward. The position angle of the line of nodes
is $\theta = 142\arcdeg$ (measured from north to east).
We assume that the LMC's center is at $(RA, Dec.) = (81.90\arcdeg, -69.87\arcdeg)$ and the distance to the LMC 
is 50.1 kpc (vdM02). The tangential velocity  of the center of mass and its orientation is that from V10, and its
line of sight velocity is $262.1\pm3.4$ km~s$^{-1}$ (vdM02).

In Figure 18, we show the proper-motions with respect to the LMC's center of mass 
$\mu_{\perp}$ (top) and $\mu_{\parallel}$ (bottom) of the captured-SMC stars 
as a function of $\xi$ coordinate. These are from the local solution
$\# 1$ with 50 reference stars (see Tab. 2), i.e., one treatment of the six.
The shaded areas represent the 1-sigma error of linear fits to the data.
We also show the rotation curve of the LMC disk (red line), the rotation curve of a 
counter-rotating disk (dashed red line)
(i.e., in the same plane as the LMC disk, and with the same amplitude rotation, but in the opposite sense), 
and that of a prograde disk with zero inclination angle, and amplitude of 200 km~s$^{-1}$ (blue line).
In the top plot, these models are easily distinguishable, however, in the bottom
plot they depart slightly from zero, and nearly coincide, indicating that rotation has a very small
contribution to $\mu_{\parallel}$.

In Figure 19 we show a zoomed-in version of Fig. 18, where  
the data points are replaced with the error-weighted average values of the proper motions
at positive and negative $\xi$. Each weighted average value is from one of the six different 
local-type solutions (Tab. 2). The $\pm1\sigma$ range of the $\xi$ coordinate at
positive and negative $\xi$ is shown 
with grey lines. The representation of the three model disks is the same as in Fig. 18.

The two solutions proposed by O11 for the captured-SMC stars are as follows.
The first is a disk of similar inclination to that of the LMC, 
specifically $20\arcdeg\pm3\arcdeg$, but rotating in the opposite sense, with the line of nodes
oriented at $\theta = 177\arcdeg\pm7\arcdeg$. They refer to this solution as the counter-rotating disk.
Although our model for the counter-rotating disk differs slightly from the O11 one, namely
by $\sim 15\arcdeg$ in inclination, and $\sim 35\arcdeg$ in the orientation of the line of nodes, these 
slight changes do not affect our conclusions. For instance, the rotation velocity is projected onto 
$\mu_{\perp}$ as $V_{rot}\times cos~i$, where $i$ is the inclination angle. A change of $i=34.7\arcdeg$
to $20\arcdeg$ implies a change of $\sim 10\%$ in $V_{rot}\times cos~i$, or of the order of $0.04$ mas~yr$^{-1}$.
A change of $35\arcdeg$ in the orientation of the line of nodes, implies a change of $18\%$ in 
$V_{rot}\times cos~i$, or $0.05$ mas~yr$^{-1}$. These values are well below our uncertainties.

From Fig. 19, top panel, it is clear that
the counter-rotating disk model is disfavored by the observations, in fact it is discrepant at a
$4.9\sigma$ level.
The second solution proposed by O11, is a disk inclined from the plane of the sky
at $-19\arcdeg\pm2\arcdeg$, with a similar orientation of the line of nodes as in their first solution, and
rotating in the same sense as the LMC disk. Our proper-motions support this 
solution, as the sense of rotation is the correct one: positive slope in the run
of $\mu_{\perp}$ with $\xi$. The proper-motion amplitude should be larger than for 
prograde rotation with $i = 34.7\arcdeg$, since the inclination angle is smaller in absolute value,
but only by $\sim 0.04$ mas~yr$^{-1}$ (see above).
The limiting case of a disk with $i=0\arcdeg$ and amplitude $V_{rot} = 200$ km~s$^{-1}$
is shown with a blue line. The proper motions imply prograde rotation in excess of  200 km~s$^{-1}$.
However, it is very likely that the motions of captured-SMC stars are non-circular. 
That this may be the case, is also implied by the run of $\mu_{\parallel}$ with $\xi$ (Fig. 19, bottom).
The observations disagree with all models at a $\sim 2\sigma$ level.

The escape velocity from the LMC at a distance of 2.5 kpc from its center ($\xi \sim 2.9\arcdeg$, from Fig. 19)
is  $V_{esc} = 148$ km~s$^{-1}$.  This is estimated by adopting
a Plummer potential for the LMC with a total mass of $10^{10}$ M$_{\odot}$, and a scale radius of 3 kpc
(DB12).  This is a rather low mass for the LMC; a factor of two larger mass would provide
an escape velocity of 210 km~s$^{-1}$.
Our proper motions thus indicate that the captured-SMC stars 
are near the upper limit of velocities for a bound status.

\begin{table*}[htb]
\caption{Proper-motion Results}
\begin{tabular}{rll}
\tableline
\\
 \multicolumn{1}{c}{Solution}  &  \multicolumn{1}{c}{$\Delta \mu_{\alpha}$} &  \multicolumn{1}{c}{$\Delta \mu_{\delta}$} \\
\tableline \\
 \multicolumn{3}{c}{local \#1, ave.}   \\ 
\multicolumn{1}{r}{$200$,box20} &  $-0.85(0.53)-[-0.20(0.11)] = -0.65\pm0.54$ & $-0.01(0.50)-[-0.06(0.09)] = +0.04\pm0.51$  \\ 
\multicolumn{1}{r}{$50$,box20} & $-0.43(0.54)-[-0.13(0.11)] = -0.30\pm0.55$ & $-0.10(0.50)-[-0.07(0.09)] = -0.03\pm0.51$ \\ 
\multicolumn{1}{r}{$200$,box40} & $-1.37(0.70)-[-0.20(0.19)] = -1.17\pm0.72$ & $-0.64(0.69)-[-0.21(0.09)] = -0.43\pm0.69$ \\ 
\multicolumn{1}{r}{$50$,box40} & $-1.12(0.70)-[-0.09(0.19)] = -1.03\pm0.70$ &  $-0.69(0.69)-[-0.28(0.11)] = -0.41\pm0.70$ \\ \\
 \multicolumn{3}{c}{local \#1, w.a.}   \\ 
\multicolumn{1}{r}{$200$,box20} & $-0.87(0.27)-[-0.08(0.13)] = -0.79\pm0.30$ & $ +0.60(0.27)-[+0.55(0.11)] = +0.05\pm0.29$ \\ 
\multicolumn{1}{r}{$50$,box20} & $-0.35(0.29)-[+0.01(0.13)]  = -0.36\pm0.32$ &  $+0.53(0.28)-[+0.44(0.11)] = +0.09\pm0.30$  \\ 
\multicolumn{1}{r}{$200$,box40} & $-0.58(0.25)-[-0.02(0.16)] = -0.56\pm0.30$&  $-0.07(0.25)-[+0.46(0.09)] =-0.53\pm0.26$ \\ 
\multicolumn{1}{r}{$50$,box40} & $-0.24(0.27)-[+0.08(0.15)]  = -0.32\pm0.31 $ & $ -0.03(0.27)-[+0.32(0.11)] =-0.35\pm0.29 $\\ \\ 
 \multicolumn{3}{c}{local2, ave.}   \\
\multicolumn{1}{r}{$200$,box20} & $ -0.45(0.53)-[-0.15(0.11)] = -0.30\pm0.54 $ & $ +0.14(0.49)-[-0.13(0.10)] = +0.27\pm0.50$ \\ 
\multicolumn{1}{r}{$50$,box20} & $-0.18(0.53)-[-0.21(0.10)] =  +0.03\pm0.54 $ & $-0.03(0.50)-[-0.17(0.10)] = +0.14\pm0.51$ \\ 
\multicolumn{1}{r}{$200$,box40} & $-1.21(0.70)-[-0.02(0.18)]  = -1.19\pm0.72$ & $ -0.22(0.66)-[-0.25(0.09)] =  +0.03\pm0.67$ \\ 
\multicolumn{1}{r}{$50$,box40} &$ -1.02(0.70)-[-0.16(0.19)] = -0.86\pm0.72$ & $-0.75(0.70)-[-0.28(0.11)] = -0.47\pm0.71$\\ \\
 \multicolumn{3}{c}{local2, w.a.}  \\
\multicolumn{1}{r}{$200$,box20} & $-0.42(0.27)-[-0.02(0.12)] = -0.40\pm0.29$ & $ +0.62(0.27)-[+0.34(0.10)]  =  +0.28\pm0.29$ \\ 
\multicolumn{1}{r}{$50$,box20} &$ -0.22(0.28)-[+0.01(0.12)]  = -0.23\pm0.30  $  & $ +0.36(0.29)-[+0.16(0.09)]  =  +0.20\pm0.30$ \\ 
\multicolumn{1}{r}{$200$,box40} & $ -0.24(0.26)-[+0.16(0.15)]  = -0.40\pm0.30$ & $  +0.16(0.26)-[+0.28(0.11)]  = -0.12\pm0.28$ \\ 
\multicolumn{1}{r}{$50$,box40} & $-0.07(0.27)-[+0.12(0.15)]  = -0.19\pm0.31 $ & $ -0.05(0.27)-[+0.09(0.10)]  = -0.14\pm0.29$ \\ \\
 \multicolumn{3}{c}{local3, ave.}   \\
\multicolumn{1}{r}{$200$,box20} & $-0.45(0.54)-[-0.15(0.08)] = -0.30\pm0.54 $ & $  +0.21(0.50)-[-0.01(0.10)] = +0.22\pm0.51$ \\ 
\multicolumn{1}{r}{$50$,box20} & $-0.06(0.51)-[-0.16(0.10)] =  +0.10\pm0.52$ & $ +0.32(0.50)-[-0.08(0.10)] =  +0.40\pm0.51$ \\ 
\multicolumn{1}{r}{$200$,box40} & $-1.15(0.72)-[-0.10(0.17)] = -1.05\pm0.74 $ & $-0.31(0.67)-[-0.11(0.11)] = -0.20\pm0.68$ \\ 
\multicolumn{1}{r}{$50$,box40} & $-1.13(0.70)-[-0.12(0.19)] = -1.01\pm0.72$ & $ -0.26(0.69)-[-0.25(0.11)] = -0.01\pm0.70$ \\ \\
 \multicolumn{3}{c}{local3, w.a.}  \\
\multicolumn{1}{r}{$200$,box20} & $-0.48(0.28)-[-0.01(0.12)] = -0.47\pm0.30$  & $ +0.55(0.28)-[+0.39(0.11)]  = +0.16\pm0.30$ \\ 
\multicolumn{1}{r}{$50$,box20} & $ +0.00(0.29)-[+0.02(0.13)]  = -0.02\pm0.32$ & $ +0.52(0.29)-[+0.22(0.09)]  = +0.30\pm0.30$ \\ 
\multicolumn{1}{r}{$200$,box40} & $ -0.29(0.26)-[+0.15(0.16)]  = -0.44\pm0.30$ & $  +0.09(0.26)-[+0.34(0.11)]  = -0.25\pm0.28$ \\ 
\multicolumn{1}{r}{$50$,box40} & $ -0.09(0.27)-[+0.15(0.15)]  = -0.24\pm0.31 $ & $ +0.14(0.27)-[+0.12(0.10)]  = +0.02\pm0.29$ \\ 
\tableline
\end{tabular}
\end{table*}

\begin{figure}
\includegraphics[scale=0.5,angle=-90]{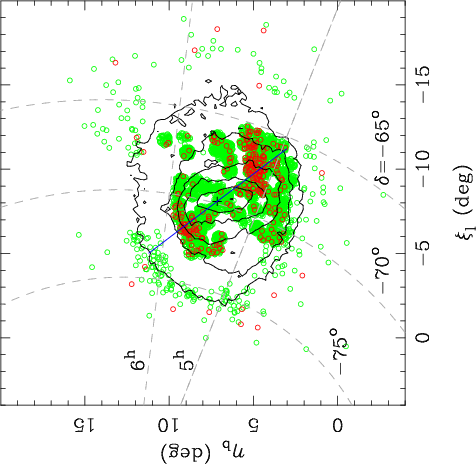}
\caption{Spatial distribution of the captured-SMC sample from O11 (red), and
bona fide LMC sample (green). M-giant density contours are shown in black. 
The coordinate system is the same as in Fig. 8-10. The LMC's center is marked with a cross.
The line of nodes of the LMC disk is shown with a blue line.}
\end{figure}

\begin{figure}
\includegraphics[scale=0.4,angle=-90]{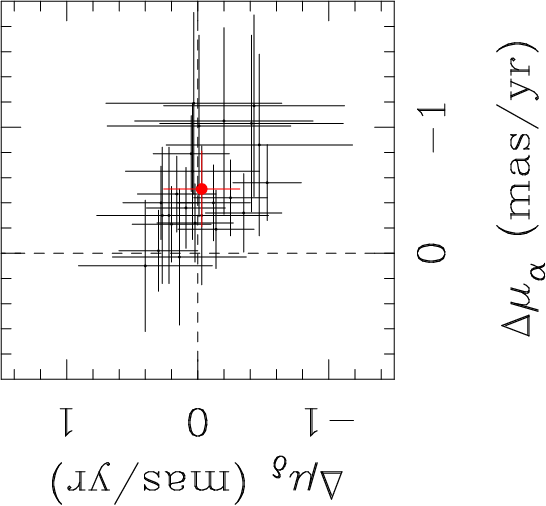}
\caption{Twenty-four determinations of the proper-motion difference between the
captured-SMC sample and the LMC sample; the average value of these determinations is shown in red.}
\end{figure}

\begin{figure}
\includegraphics[scale=0.48]{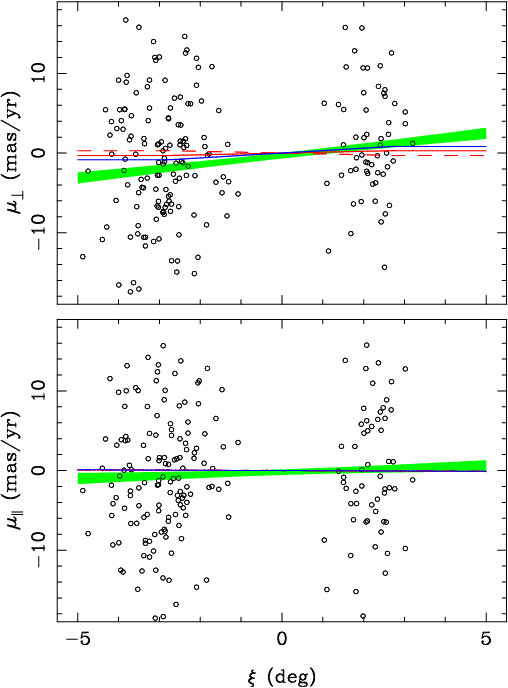}
\caption{Proper motions perpendicular (top) and parallel (bottom) to the line of nodes, as a function of
position along the line of nodes (open symbols). Proper motions are with respect to the LMC's center of mass
proper motion.
The green shaded areas show the $1\sigma$-error of a linear fit to the data. 
The rotation of the LMC disk is shown with a red line; that of a counter-rotating disk
with a dashed red line. A positive slope in $\mu_{\perp}$ indicates prograde rotation. 
The blue line indicates prograde rotation
with zero inclination angle, and amplitude $V_{rot} = 200$ km~s$^{-1}$. In the bottom plot, 
these three models nearly coincide. The data points represent one of the six types of 
solutions presented in Tab. 2.}
\end{figure}

\begin{figure}
\includegraphics[scale=0.48,angle=0]{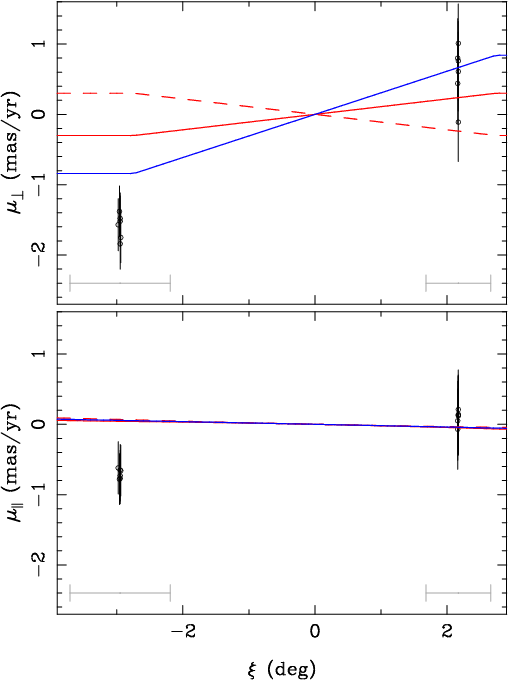}
\caption{The data points represent weighted averages for the six types
of local solutions (see Tab. 2 and text) at positive and negative $\xi$. 
The rotation curve of the LMC disk is shown in red, that of a 
counter-rotating disk in dashed red, and that of a prograde disk with zero inclination and
amplitude $V_{rot} = 200$ km~s$^{-1}$ in blue. The $\pm 1\sigma$ $\xi$ range of the
dta is also shown with grey lines.}
\end{figure}

\section{Discussion}

\subsection{Implications of the Spatial Distribution of OB Candidates}

Our sample of OB candidates was purposely built with rather restrictive constraints:
we aimed to produce a reliable map of the spatial distribution
over an expansive area, rather than a complete sample of candidates. For instance,
the comparison with 6X 2MASS indicates that our errors in $J$ magnitudes, as given by 2MASS
at the faint end, tend to overlook candidates rather than contaminate our sample with non-candidates. 
The color cuts also were designed to secure the bluest, i.e., youngest objects 
in our sample, and to eliminate foreground subdwarfs which are more frequent
at B-type than O-type (Vennes et al. 2011 and references therein).
These types of stars have been studied starting in the early nineties, with
the photographic work by Irwin et al. (1990) in two UK Schmidt fields located between the Clouds.
Deep, follow-up CCD photometry on small subfields selected from the Irwin et al. (1990) 
study by Demers \& Battinelli (1998) indicated the existence of associations
with ages as young as 10-25 Myr. Our work supports this notion, although
ages as old as $\sim 90$ Myr are allowed by our color cuts 
(main-sequence lifetime of a B5 type star). 
The notion that very young stars ($\sim 10-30$ Myr) are present in the eastern part of the
SMC wing is supported by the finding of three high mass X-ray binaries with 
optical counterparts of confirmed 
(Kahabka \& Hilker 2005) and implied (McBride et al. 2010)
Be spectral type. A 10 Myr-lived  star located some 7 kpc ($6.4\arcdeg$) from the SMC's center
requires an excess velocity of 700 km~s$^{-1}$ to travel this distance.
Kahabka \& Hilker (2005) measure the radial velocity of their Be optical
counterpart to be no different than 30 km~s$^{-1}$ from the SMC's velocity.
Therefore, these young stars must have formed in situ rather than forming in the 
inner regions of the SMC, and migrating to their current location.

Most remarkably, the LMC periphery shows two stellar concentrations at the ends of the bar,
and  a well populated distribution in the north-east region.
In some nearby barred spiral galaxies (e.g., Sheth et al. 2002), 
star-formation activity appears enhanced at the 
ends of the bar, indicating perhaps the beginning of rings. 
Our Milky Way too has been shown to have young and massive
(ages $\le 50$ Myr and masses $\ge 10^{4}$ M$_{\odot}$)
stellar clusters  at the near and far ends of the bar (Davies et al. 2012 and references therein).
In the case of the LMC however, we have a clear case of a recent interaction with the SMC,
which is responsible for the newly formed stars in the LMC's periphery, as well as
captured-SMC stars in the LMC. We also know that the recent star-formation history 
of the Clouds is coupled, noting the recent study of the ages of
young ($ < 1$ Gyr) clusters in the Clouds by 
Glatt et al. (2010). Their study indicates that
there were two periods of enhanced star formation in both Clouds, and that these are nearly
coincident: in the LMC at 125 Myr and 800 Myr, and in the SMC at 160 Myr and 630 Myr. Clearly,
the recent epoch of enhanced star formation in both Clouds is due to the recent interaction implied by the orbits.

For the first time, we present the prospect of finding newly formed stars in the LA.
The existence of two OB-candidate clumps in regions of high H I column density in the LA
needs to be further explored spectroscopically for confirmation. The LA
is the most challenging structure to reproduce in models of the Magellanic system, and as 
discussed by previous studies (e.g., DB12), the evolution of this material in the 
recent past (300 Myr) should be influenced by hydrodynamical interactions with the MW hot gaseous halo
(i.e., ram pressure). 

\subsection{Kinematical Evidence Regarding the Most Recent Interaction of the Clouds}

In order to explore their recent interaction, we use the proper motions of the Clouds as
measured by K06 and V10 to integrate their orbits back in time as individual 
point masses in a realistic, three-component Milky-Way gravitational potential (Johnston et al. 1995).
We then calculate the minimum separation between the Clouds (impact parameter), the time as well as
the relative velocity at minimum separation. We also calculate the 
angle between the relative velocity and the LMC disk at minimum separation.
We assume the LMC's disk has an inclination of $34.7\arcdeg$ and the position angle of the line of 
nodes is $\theta = 142\arcdeg$ (e.g., O11). The present disk inclination corresponds to an
orientation in Galactocentric coordinates that is assumed not to change with time.
As we are interested in the most recent interaction between the Clouds,
we integrate back in time for only 400 Myr, compared to several Gyr-period orbits (or unbound orbits,
Besla et al. 2007).

We vary the input proper motions according to their formal errors, (where for V10 we use
the errors as obtained in the Clouds' relative proper-motion estimation, i.e., $0.11$ mas~yr$^{-1}$
per Cloud). Distances, radial velocities and the Clouds' locations are taken from V10. 
Proper-motion errors are assumed to have a Gaussian distribution; 
the distances and radial velocities of the Clouds 
are assumed to have zero errors. We integrate 1000 such realizations, and determine the range for 
the inner $68.2\%$ of the values. Results are summarized in Table 3, where the first column indicates the 
reference to the proper-motion measurement, the second column shows the minimum separation and the third shows 
the time when the encounter occured. The fourth column shows the relative velocity at minimim separation, and
the fifth column shows the angle between the relative velocity and the LMC disk, at minimum separation.
\begin{table}[htb]
\caption{Parameters of the Most Recent  Interaction}
\begin{tabular}{rrrrr}
\tableline
\\
  & \multicolumn{1}{c}{d$_{min}$ (kpc)} & \multicolumn{1}{c}{t (Myr)} & \multicolumn{1}{c}{V$_{rel}$ km~s$^{-1}$}  & \multicolumn{1}{c}{Angle (deg)}\\
\tableline
V10 & $11.6^{+6.2}_{-2.1}$ & $223^{+35}_{-108}$   &  $91.6^{+54.6}_{-22.3}$  &  $125.3^{+4.7}_{-10.3}$   \\
K06 & $13.9^{+8.9}_{-4.7}$ & $234^{+20}_{-176}$   &  $82.4^{+80.8}_{-15.4}$  &  $110.5^{+10.5}_{-35.5}$   \\
\tableline
\end{tabular}
\end{table}
We note that in the K06 tests, 47 realizations place the minimum separation in the future,
i.e., the Clouds are currently approaching one another. Also, the distribution of the minimum separation is rather 
flat between 9 and 20 kpc, and increases toward larger minimum separations.
In the V10 tests, there are only 8 such occurrences, and the distribution of the minimum 
separation has a peak at $\sim 12$ kpc; overall the V10 parameters are better constrained than those using K06. 
This simple approach neglects dynamical friction of the Clouds with the Milky Way and the gravitational interaction
between the Clouds. Since the tidal radius of the LMC is $\sim 15$ kpc, the gravitational effect of the LMC on the SMC
will become important for separations smaller than this value. As such, our estimates of the
minimum separation are to be regarded as upper limits. Our simple approach indicates that the 
impact angle of the SMC with respect to the LMC disk is
$35\arcdeg^{+5}_{-10}$ away from being coplanar according to the V10 proper motions, and smaller assuming K06 (Table 3).
That the impact is not coplanar is also implied by our analysis of the O11 captured-SMC stars, which favors 
the interpretation that thses  to
move in a plane inclined by  $\sim 54\arcdeg$ from LMC's disk.

To fully understand this interaction, N-body studies of both Clouds in a realistic MW potential are needed.
One such recent study is that of DB12, where the interaction is modeled after a preliminary 
thorough search in the orbital parameter space that was centered on the V10 proper motions. 
Their search for a realistic model is further constrained by key observational facts such as the 
recent ($\le 2$ Gyr) formation of the binary
pair necessary for the formation of the MS, the actual spatial distribution and kinematics of the MS, and the current 
positions and velocities of the Clouds. The only drawback of this major study that aims to 
reproduce the MS, Bridge, and SMC morphology, is that the  
LMC's morphology was not captured, nor was it meant to be, in accordance with the initial goals of the study; that is
the SMC is described by an N-body system, while the LMC and MW have analytic potentials.
The best model of DB12 places the most recent interaction at a minimum separation of 6.6 kpc some $t=260$ Myr ago.
We also mention the recent
work by Besla et al. (2012) who use an N-body smoothed-particle hydrodynamics (SPH) code 
to model the interaction, where both Clouds are modeled as N-body/SPH systems with disks, 
and the MW has a static potential.
Besla et al. (2012) focus on describing the particular morphology of the LMC in light of a collision 
between the Clouds that took place in the recent past. The LMC model has a bar before the interactions take place.
However, their modeling does not reproduce the
current position and velocity of the SMC.
The model is constrained by the orbit of the LMC as measured by K06, and
other facts that point to a recent interaction between the Clouds.
Among these are: the coupled recent star-formation history of the Clouds (e.g., Harris \& Zaritsky 2009), 
the off-center stellar bar in the LMC (the center of the bar is offset by $\sim 1.2\arcdeg$ from the
kinematic center of H I; vdM02),
the Bridge and its gaseous and stellar makeup, and the existence of captured-SMC stars in the LMC (O11).
Their description of the interaction accounts for many features of the particular morphology of the LMC,
with the most recent interaction having occured  100 Myr ago and being a central collision at high inclination 
with respect to the LMC's disk.

The questions that remain to be answered are related to the particulars of this interaction, such as
impact parameter and impact angle with respect to the LMC's disk.
These will determine many of the properties of the present day LMC, SMC and Bridge.
We qualitatively explore the N-body/SPH models by 
Berentzen et  al. (2003) that describe the interaction between a barred disk galaxy
and a small ($20\%$ mass of the disk galaxy) spherical perturber.
The gas amounts to $30\%$ of the disk mass. These models analyze 
nearly perpendicular passages with respect to the disk plane, and three impact parameters: central
(0 kpc), minor-axis impact (3 kpc) and major-axis impact (6 kpc). All these interactions produce
expanding ring structures, off-center bars, spokes and other asymmetries in the stars and gas,
and are unlike coplanar interactions that primarily influence the strength of the existing bar
of the host galaxy (Berentzen et al. 2004). Among the three classes of models
presented by Berentzen et  al. (2003), the central and minor-axis impact have their stellar bar destroyed
in $\sim 200$ Myr. This is not the case of the LMC, which has a strong bar seen in its M-giant population.
In the major-axis impact, the stellar bar survives, has its maximum displacement (2.4 kpc) from
the initial position 100 Myr after impact, and recenters after 600 Myr (some two disc rotations).
The impact produces a radially expanding density wave which becomes first visible in the gas.
Since the impact is off-center, the symmetry of the gaseous ring is broken when it encounters the stellar bar
or spiral arms. Nevertheless, gaseous density enhancements in ring-like shapes and 
at the ends of the bar appear to form (see Fig. 17 and 22 in Berentzen et  al. 2003, see
also simulations by Bekki \& Chiba 2007, their Figs 4 and 5).
Thus, the distribution of the OB candidates in the LMC's periphery, the timing of the interaction,
and the characteristics of the bar point to an off-center collision of moderate to high inclination 
with respect to the host's disk.  
Although the $1\sigma$ lower limit of the impact parameter is $\sim 9$ kpc from our
simple analysis of the proper-motion data, this may not represent a problem 
when more realistic models are considered (e.g., DB12).

The recent interaction of the Clouds also tidally strips material from the SMC disk in the form
of gas and stars and forms the observed Magellanic Bridge. This is clearly seen in the recent work of
Besla et al. (2012) and DB12, as well as in some earlier work.
For the reasons mentioned above, we follow in more detail the prediction of the
DB12 work. They predict that the interaction at $\sim 200$ Myr ago
produced two tidal structures, one filling the region between the Clouds, and representing
the observed Bridge, and the other --- the Counter Bridge --- 
extending away from the SMC, but rather aligned  with the line of 
sight to the SMC, and thus hidden behind the SMC. 
The simulations show that the Counter Bridge has an extent of $\sim 20$ kpc. It is
known that the SMC has a significant depth along the line of sight $12-14$ kpc 
(Subramanian \& Subramanian 2012, Crowl et al. 2001). Here we suggest the possibility
that the two branches of OB candidates extending from the western part of the SMC wing and surrounding
the Cloud are part of this Counter-Bridge structure. Spectroscopic information to secure 
distances and radial velocities is necessary to further explore this matter.
The DB12 model also predicts the existence of tidally stripped stars in the Bridge.
We note that the photometric study  by Harris (2007) that sampled twelve fields
in the Bridge, found no older stars. However, his fields were placed along the
ridge line of the H I density, which appears to be offset from the OB candidates.
One field, ($\# 14$, (RA, Dec.) = ($03^h18^m, -73\arcdeg 18\arcmin$) )
from his study was placed $\sim 1\arcdeg$ north 
from the H I  ridge line, i.e., closer to the location of our OB candidates.
Monelli et al. (2011) reinvestigate two of the Harris (2007) fields, including
field $\#14$ with deep and wide-field photometry. Their Fig. 3 shows a well-defined main sequence
in field $\#14$ as well as in  field $\#13$, thus reinforcing the tidal nature of the Bridge.

\section{Summary}

Based on photometry from GALEX, 2MASS, SPM4 and APASS and on
proper motions from SPM4, we select a sample of distant (low proper-motion) OB-type candidates
in a 7900 square degree area, for the purpose of mapping recently formed stars
in the Magellanic Clouds' periphery and  Bridge,
and to explore the presence of such stars in the Leading Arm and Magellanic Stream.
We find a well populated LMC periphery with two prominent 
clumps at the ends of the LMC bar, and the northern side being
more abundant than the south. In agreement with earlier work,
we also find a  rich SMC wing that continues eastward in a narrow band 
toward the LMC to RA $\sim 3.4^h$. This path is offset by $1\arcdeg-2\arcdeg$ from the
H I ridge line in the Bridge. A group of 2-3 candidates is present in the LMC periphery at
R.A. $= 4.2^h$ in the direction of the band extending from the SMC wing, and
suggesting that this is the site where material from the SMC is funneled into the LMC.
The western part of the SMC wing splits into two branches that partly surround the SMC;
they may be part of the Counter Bridge tidal structure produced in the most 
recent interaction between the Clouds as modeled by DB12.

We also find clumps of OB candidates in the LA and coincident with high H I 
density regions. Although these clumps are located at low Galactic latitudes,
the known spiral structure of the MW does not offer a compelling alternative explanation.
A few isolated OB candidates are also seen in the MS. Spectroscopic investigation of
these candidates is most desirable.

Simple calculations based on the Clouds' proper motions suggest that
the most recent interaction took place some $200$ Myr ago, with an impact parameter
between 9 and 18 kpc (from the V10 proper motions). 
The sophisticated modeling by DB12 indicates a similar impact time,
with an impact parameter of 6.6 kpc. The orbits of the Clouds,  taken together with aspects such as
the off-center LMC bar, the captured SMC stars in the LMC (O11), and the configuration of the
OB candidates in the LMC, imply that this interaction was an off-center, moderate to high-inclination
collision between the SMC and LMC's disk.

We also analyze the SPM4 proper motions of the radial-velocity sample from 
O11 of $\sim 5900$ red supergiants, 
oxygen- and carbon-rich asymptotic giant branch stars and other giants.
O11 showed that $5\%$ of these stars have distinct kinematics from the rotation of the LMC disk, and
proposed that these stars are captured from the SMC. O11 also show that the metallicities of 30 such stars 
are compatible with the metallicity of the SMC's outer regions. Under the assumption of circular 
motions, O11 propose two kinematical solutions for the captured-SMC stars:  a counter-rotating disk,
i.e., similar plane with that of the LMC disk, and retrograde rotation, and a prograde rotation of
a disk some $54\arcdeg$ away from the LMC disk plane, i.e., at $-19\arcdeg$ inclination with the plane of the sky.
The proper motions indicate that 1) the captured-SMC group is kinematically distinct from the LMC disk
sample, 2) the counter-rotating disk solution proposed by O11 on radial-velocities alone 
is rejected at $ > 4 \sigma$ level, while their second solution of a prograde rotation at inclination $-19\arcdeg$ 
appears to be a compatible description, although the assumption of circular motions 
may be unrealistic, and 3) the implied orbital velocities for the captured-SMC 
star indicate a  marginally bound status to the LMC.
Our proper-motion analysis thus reinforces the hypothesis put forward by O11, 
that these are not bona fide LMC stars, but rather captured from the SMC during 
their close interaction in the recent past (e.g., DB12 and references therein).
%

We acknowledge support by the NSF through grant AST04-0908996.
We thank Stephane Vennes for making available the model colors for
main sequence and white dwarf stars. 
We are grateful to Knut Olsen for providing the radial-velocity catalog
in the LMC, and to Mathew Templeton from AAVSO for making available the
APASS-DR3 catalog. We thank David Nidever for making available the H I column 
density data and for helpful discussions regarding this study.
This publication makes use of data products from the Two Micron All Sky Survey,
which is a joint project of the University of Massachusetts and the
Infrared Processing and Analysis Center/California Institute of Technology,
funded by the NASA and the NSF.
This research was made possible through the use of the 
AAVSO Photometric All-Sky Photometric Survey, funded by Robert
Martin Ayers Sciences Fund.

\end{document}